# Self-similarity of low-frequency earthquakes

M. Supino[1,2]*, N. Poiata[1,3], G. Festa[2], J.P. Vilotte[1], C. Satriano[1] and K. Obara[4]

[1]Université de Paris, Institut de physique du globe de Paris, CNRS, F-75005 Paris, France.

[2]Dipartimento di Fisica 'Ettore Pancini', Università di Napoli Federico II, I-80126, Napoli, Italy.

[3]National Institute for Earth Physics, 12 Călugăreni, Măgurele, 077125 Ilfov, Romania.

[4]Earthquake Research Institute, University of Tokyo, Bunkyo, Tokyo 113-0032, Japan.

*Corresponding author: Mariano Supino

E-mail : supino@ipgp.fr

**Low-frequency earthquakes are a particular class of slow earthquakes that provide a unique source of information on the mechanical properties of a subduction zone during the preparation of large earthquakes. Despite increasing detection of these events in recent years, their source mechanisms are still poorly characterised, and the relation between their magnitude and size remains controversial. Here, we present the source characterisation of more than 10,000 low-frequency earthquakes that occurred during tremor sequences in 2012-2016 along the Nankai subduction zone in western Shikoku, Japan. We show that the seismic moment versus corner frequency scaling for these events is compatible with an inverse of the cube law, as widely observed for regular earthquakes. Our result is thus consistent with shear rupture as the source mechanism for low-frequency earthquakes, and suggests that they obey to a similar physics of regular earthquakes, with self-similar rupture process and constant stress drop. Furthermore, when investigating the dependence of the stress drop value on the rupture speed, we found that low-frequency earthquakes might propagate at lower rupture velocity than regular earthquakes, releasing smaller stress drop.**

Worldwide, seismic and geodetic observations recorded along a number of subduction zones[1-5] and continental faults[6-8] have revealed a broad class of transient energy-release signals known as slow earthquakes. Geodetic slow earthquakes[9-12] are slow slip events (SSEs) with durations of days (short-term SSEs) or months to years (long-term SSEs). Seismic slow earthquakes are characterised by lower dominant frequencies than regular earthquakes of the same moment. These are impulsive, low-frequency earthquakes (LFEs) and tectonic tremor signals with dominant frequencies in the 1-10 Hz band[13-16], and very-low-frequency earthquake (VLFE) signals with dominant periods in the 10 s to 100 s band[17-20].

Numerous observations have shown that tectonic tremors, LFEs, VLFEs and SSEs often accompany each other and occur in ductile-to-brittle environments at the neighbouring sides of large earthquake seismogenic zones[21]. Recent observations have suggested that SSEs might trigger megathrust earthquakes[22,23]. As such, detailed characterisation of slow earthquakes activity might represent a unique source of information to improve seismic hazard monitoring and risk assessment[21]. It is often assumed that slow earthquakes provide sparse observations that probe different scales of a common transient process along slowly driven plate boundaries. While this physical process remains to be fully understood, a linear scaling between moment and source duration across the different slow earthquake observation scales has been proposed[24] and interpreted as the signature of a different process to that for regular earthquakes, or alternatively as the signature of a scale-bound source process for the longest duration events[25].

Low-frequency earthquakes are often observed in association with SSEs on the deep extensions of plate boundaries. They often occur in burst-like sequences of a multitude of events mixed in with long-lasting tectonic tremor signals. In recent years, advanced data analysis methods have been developed to improve detection of LFEs[26-28], and very large datasets are becoming available to the scientific community. However, the source mechanism and scaling properties of these events still remain poorly known, with the main difficulty being the very low signal-to-noise ratio associated with these transients. Bostock et al.[29] reported an almost constant source duration for ~100 LFE templates along the Cascadia plate boundary, over a limiting magnitude range. This result is in contrast with classical observations for regular, fast earthquakes[30], where the seismic moment is proportional to the cube of the source duration.

**The Nankai subduction zone**

Here, we present the source characterisation of 10,157 LFEs extracted from tectonic tremor sequences that occurred during the periods of May to June 2012 and January 2014 to November 2016, along the Nankai subduction zone in western Shikoku, Japan (Fig. 1).

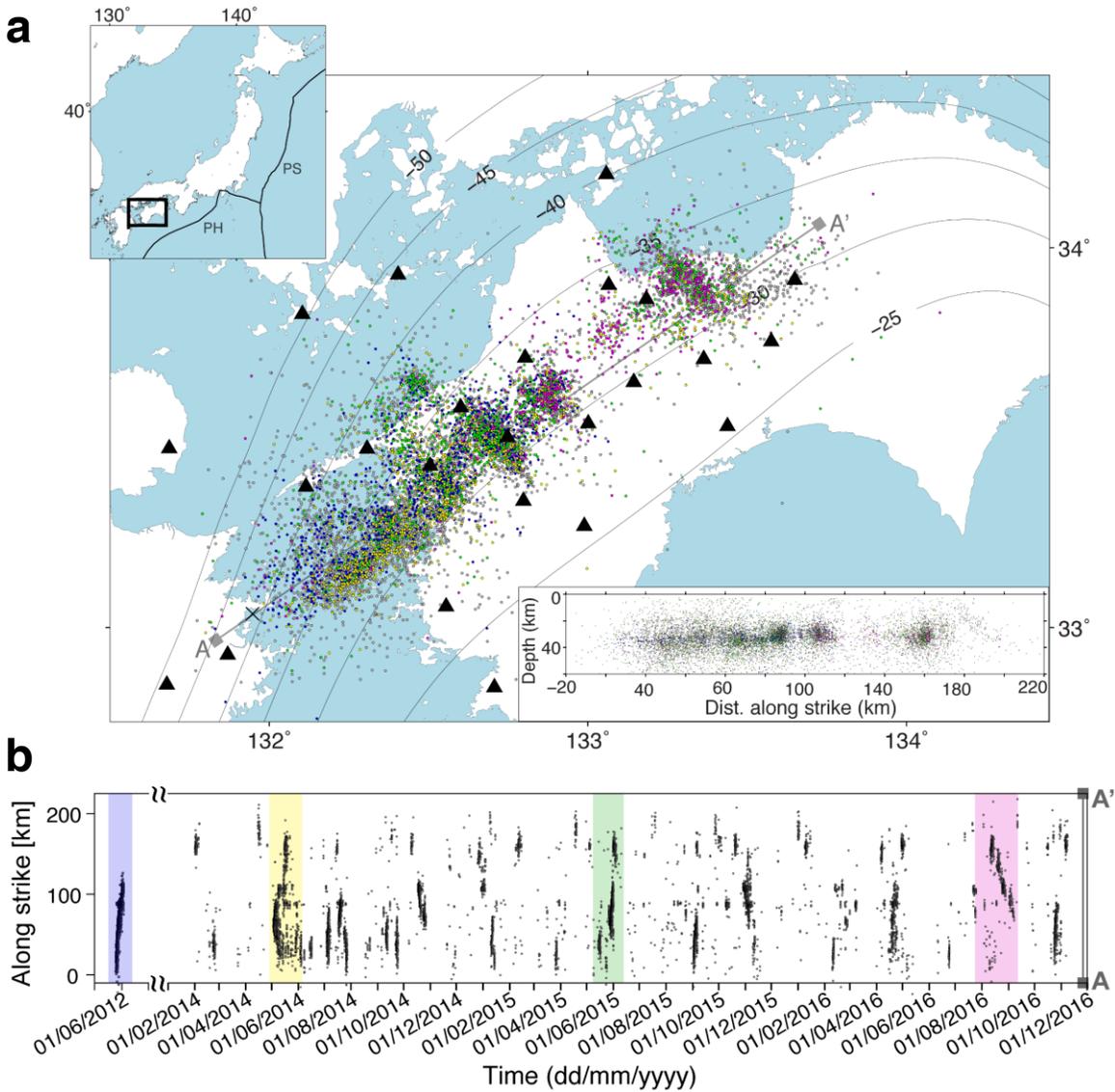

**Figure 1 | Distribution of low-frequency earthquakes. a**, Map of the locations of the analysed low-frequency earthquakes (circles) and the Hi-net stations (triangles). The event colours that are not grey indicate events extracted from the largest tectonic tremor sequence of each year analysed.

Inset, top: The geographic and tectonic settings of the western Shikoku area. Inset, bottom: The depth cross-section of events projected along the strike of N 40º E (A-A'). **b**, Space–time plot of the events. Colours as in (a).

In this region, the Philippine Sea Plate is subducting beneath Japan, with a recurrence time of megathrust earthquakes from 100 to 150 years[31]. The data analysed are velocity seismograms recorded at 25 stations of the high-sensitivity borehole seismic network (Hi-net), managed by the National Research Institute for Earth Science and Disaster Prevention (NIED), Japan[32,33]. LFEs have been detected and located by exploiting the coherency of the wavefield characteristics recorded across the network stations[28,34] during the periods of major tectonic tremor activity[35,36]. The derived LFE catalogue consists of over 40,000 events (Supplementary Fig. 1).

**Source characterisation**

We processed the data and characterised the source parameters for each of the events in the catalogue. We modelled the S-wave displacement amplitude spectrum of the LFEs using a generalised Brune's spectral model[37] (see Methods). We assumed a horizontally layered one-dimensional propagation model, and a constant frequency-independent anelastic attenuation factor $Q$, which is a good approximation for the area investigated[38].

After removal of the Green's propagator, the source spectrum is assumed to be flat at low frequencies and to decay as a power law at high frequencies, with a cross-over region around a cut-off corner frequency $f_c$. The parameters to be retrieved are: the flat spectrum level, which is proportional to the seismic moment and a proxy for the event magnitude; the corner frequency, which is related to the event size; and the power-law of the high-frequency fall-off, which constrains the energy radiated by the earthquake. The source parameters are estimated by inversion of the spectra, using a probabilistic approach[39]. This method evaluates the joint probability density functions (PDFs) of the source parameters, which allows robust estimations, accounting for between-parameter correlations in the final estimates of the parameters and related uncertainties.

For each LFE, we inverted each individual station displacement spectrum, and retrieved the marginal PDFs of each source parameter by integrating the joint PDF (see Supplementary Fig. 2; see Methods). Extremely low signal-to-noise-ratio observations were automatically detected and rejected (see Methods). The source parameters of a LFE recorded at more than one station were estimated as the weighted means of the single station solutions (see Methods), as seismic moments and corner frequencies inferred from different stations show some variability (Supplementary Fig. 3), as for regular earthquakes.

**Scaling of corner frequency with seismic moment**

The estimated corner frequency and seismic moment of the LFEs analysed are shown in Figure 2.

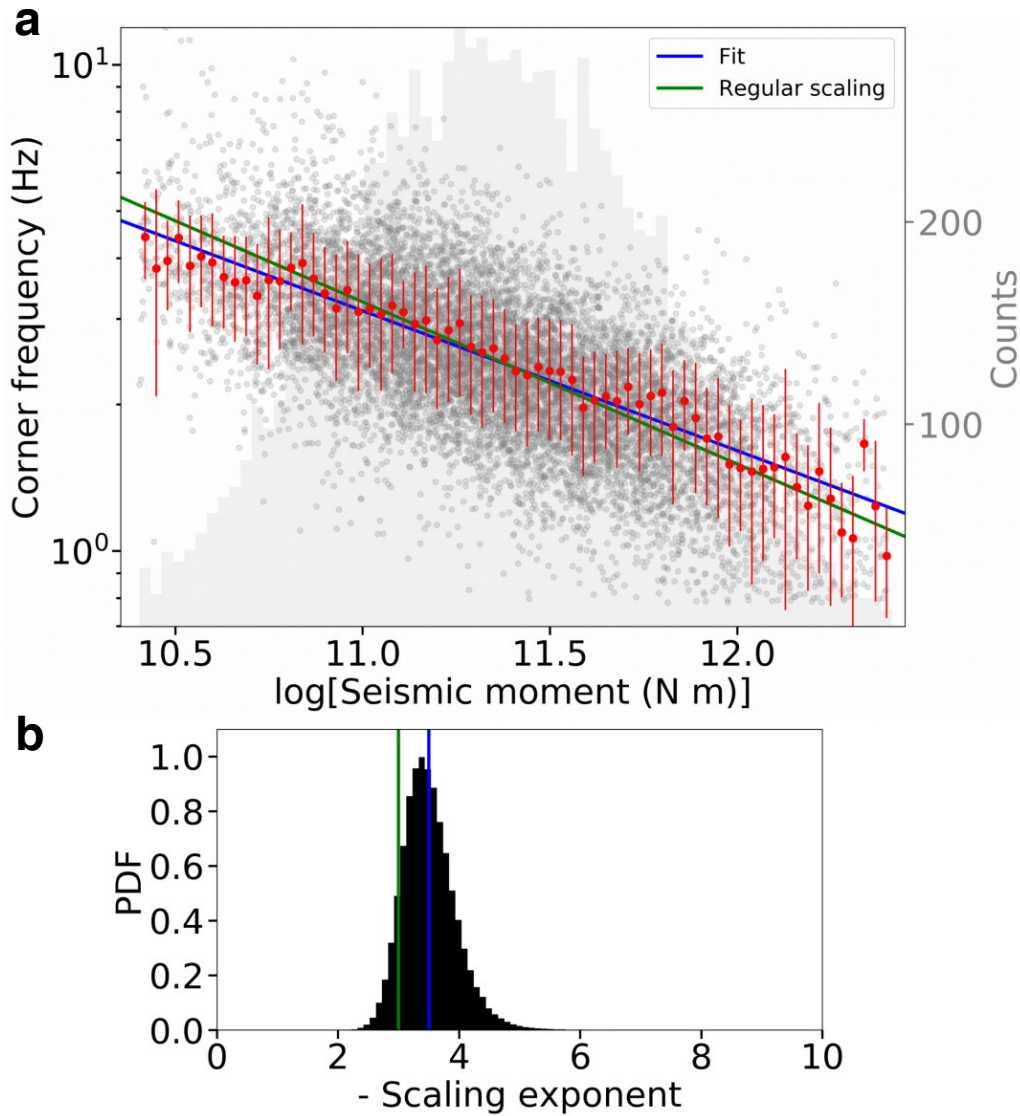

**Figure 2 | Scaling of the corner frequency with the seismic moment. a**, The best-fit curve (blue line) of the averaged estimates (red points) has a scaling exponent of -3.5. The corner frequencies and seismic moment estimates for each LFE are shown (grey points). The weighted averages of the corner frequencies for the selected seismic moment bins (bin-size, 0.03) are shown (red points), along with the weighted standard deviations per bin (red bars). The histogram in the background (grey shading) shows the number of events in each bin. The green line represents the regular scaling. **b**, Probability density function of the scaling exponent estimated with a bootstrap method performed with 100,000 random extractions (see Methods); colours as for the top panel.

The source parameters are well resolved for the whole range of the seismic moments explored. As an example, we show the velocity records, the displacement spectra and the solutions for three events from the ends and the middle of the seismic moment range explored (Supplementary Figs. 2, 4, 5). The LFEs showed typical behaviours[40,41], with corner frequencies much lower than expected for regular earthquakes of the same magnitude. The high-frequency fall-off exponents have a median of 3.0, with 80% of the events between 1.9 and 4.1.

The scaling between the corner frequency and the seismic moment is clear (Fig. 2). To deal with the large number of solutions, we grouped the corner frequency estimations into log $M_0$ bins with a size of 0.03. The histogram representing the number of points per bin is shown as the shaded background in Figure 2a. For each bin, we computed the weighted average of the corner frequencies and the uncertainty related to the data variability (Fig. 2a, red points and error bars; see Methods). Using this averaged information, we performed linear regression according to Equation (1):

$$\log f_c = A \log M_0 + B \qquad (1),$$

where $A$ and $B$ are constants to be determined (Supplementary Table 1). We obtained a scaling parameter $A' \equiv 1/A = -3.5$. We used an unweighted regression that assigns the same weight to each bin and avoids domination of the fit by the central regions in the seismic moment domain. Nevertheless, even when using weighted linear regression, the estimated scaling exponent of $A' = -3.35$ remains very close to the previous estimation.

We estimated the PDF of the scaling exponent with a Bootstrap method (Fig. 2b; see Methods), assessing the robustness of the result. We found that the scaling exponent is normally distributed, and its expected value is $-3.5 \pm 0.5$. Our result is consistent within 1-sigma confidence interval with the classical -3 scaling exponent observed for regular earthquakes[30].

We also estimated the scaling exponent when the seismic moment domain is reduced to a decade (log $M_0$ = 11.0 – 12.0), obtaining a mean value of -3.5 and a mode of -3.2 (Supplementary Fig. 6a). This test selects only the bins with at least 100 observations (Fig. 2a).

The effects of a constant anelastic attenuation factor on the $M_0$ - $f_c$ scaling was assessed by reprocessing the data with different constant attenuation factors, the lower $Q$ = 100 and the higher $Q$ = 500 than that provided in the literature (Q = 300)[38]. Results show a variation of about 15% ($-4 \pm 0.6$) for Q = 100, and of about 3% ($-3.6 \pm 0.5$) for Q = 500.

Moreover, we reprocessed the data with a frequency-dependent attenuation factor, as in Equation (2):

$$Q(f) = Q_0 \, f^{\varepsilon} \qquad (2),$$

where log $(Q_0)^{-1}$ = 2.5 and $\varepsilon$ = 0.5 as provided in literature[42]. Results (Supplementary Fig. 7) show a variation in the estimated power law exponent of about 6% ($-3.7 \pm 0.5$).

We also addressed the possibility that a -3 scaling exponent might arise when different tremor sequences from LFE clusters were collated with different scaling. We thus estimated the scaling for subsets of LFEs clustered in space and time (Supplementary Fig. 8). The LFEs were grouped by tremor sequences, and one major tremor sequence was selected per year (Fig. 1). Each cluster shows a similar scaling exponent to that derived from the entire catalogue, over almost the same seismic moment range.

Finally, we evaluated the source parameters of 1585 LFEs manually detected and located by the Japanese Meteorological Agency (JMA)[43,44], for the same region and the same time period investigated in this study. We found a scaling exponent of -3.4 (Supplementary Fig. 9).

**Regular scaling of low-frequency earthquakes**

From the LFEs analysed, we retrieved a power-law scaling between the seismic moments and the corner frequencies that is consistent with that for regular earthquakes. The seismic moment scales as the inverse of the cube of the corner frequency (Fig. 2). This is consistent with shear rupture as the source mechanism for the LFEs. Similar results have been recently reported for the scaling of SSEs in Cascadia[45] and Mexico[46], and for long-period seismicity in volcanic environments[47].

This scaling is different from that inferred by Bostock et al.[29] in the analysis of LFEs in Cascadia, where a much weaker scaling exponent between the seismic moments and corner frequencies was observed ($M_0 \propto f_c^{-10}$). It is worth to note that many LFEs were detected by Bostock et al. with a matched filter method using a template, differently from our study. Explanations other than classical shear rupture have been suggested for these latter scaling models, such as forces that act in the direction of transient fluid motion[48].

For our observations, the probability of a scaling exponent smaller than -7 is less than $3 \cdot \times 10^{-4}$ (Fig. 2b). This probability remains small (0.06) even when the seismic moment domain is reduced to half a decade (Supplementary Fig. 6b). As such, an almost flat log $M_0$ - log $f_c$ scaling is very unlikely for this dataset, even over a small range of seismic moment.

Our result also differs from the scaling exponent of ~ -1.5 that was reported for a limited number of VLFEs by Ide et al.[49], and from the scaling exponents of ~ -2.0, -2.5 estimated by Ide[50] and Ide et al.[51] using a Brownian walk model for slow earthquakes. Although, the latter 2 values are consistent within 2-sigma confidence interval with our observations when the seismic moment domain is reduced to a decade (Supplementary Fig. 6a).

Combination of the seismic moment and the rupture size allows estimation of the static stress drop[52]. The rupture size can be inferred from the corner frequency $f_c$, when a propagation model for the rupture that is governed by a specific rupture velocity $v_R$ is assumed[53]. Nevertheless, whatever the rupture model, the scaling observed implies a constant stress drop – for a constant rupture speed – and a self-similar rupture process for the low-frequency earthquakes analysed in Nankai. Although constant, the amount of the released stress drop strongly depends on the rupture speed, and it can increase by several orders of magnitude when $v_R$ is reduced. Using the circular rupture kinematic model of Sato and Hirasawa[53], we analysed the dependence of the LFE rupture size and stress drop on the rupture speed (see Methods; Supplementary Fig. 10). As shown in Figure 3, when $v_R$ decreases from $0.9\beta$ to $0.02\beta$, where $\beta$ is the shear-wave speed, the stress drop increases from $10^3$ to $10^6$ Pa.

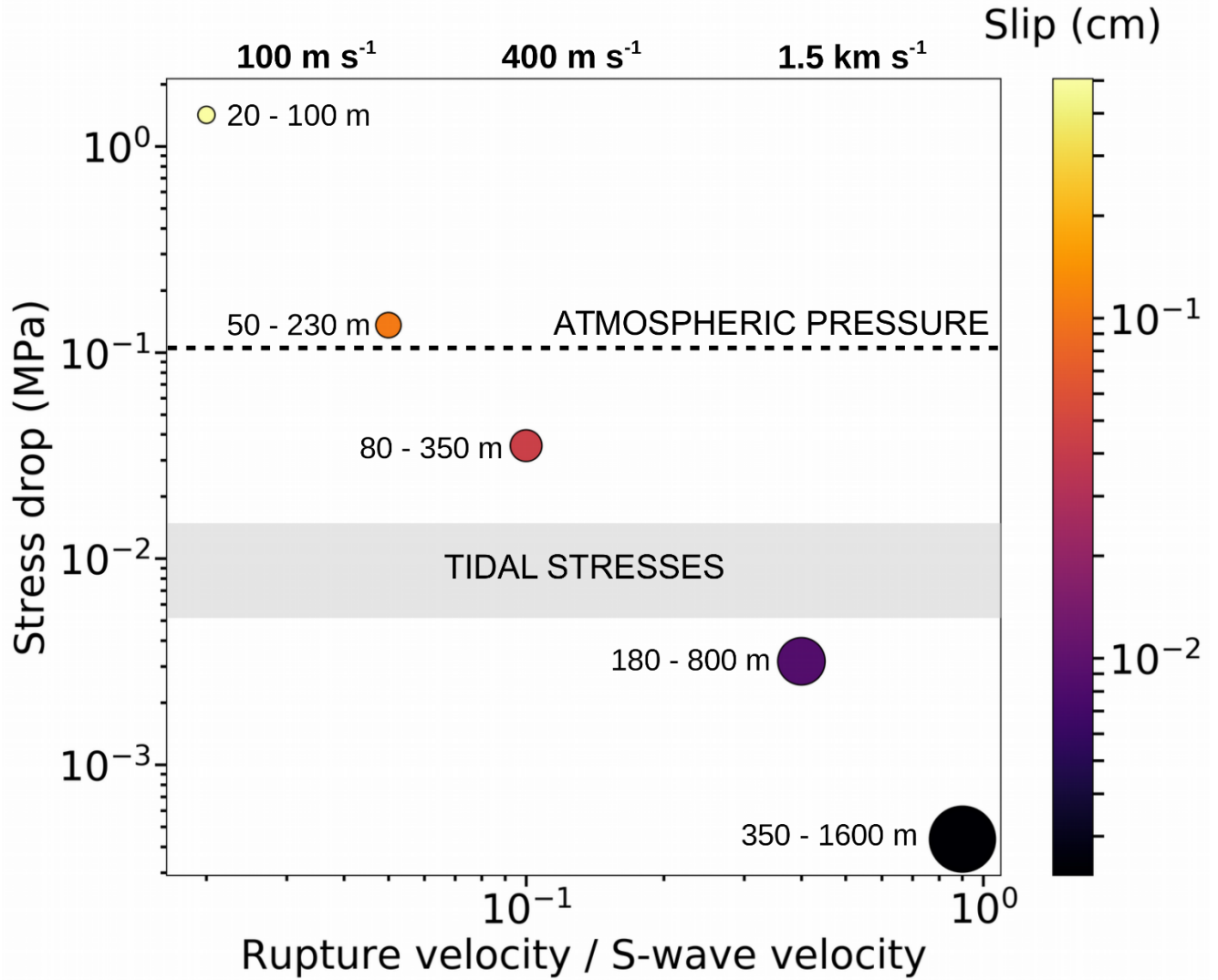

**Figure 3 | Stress drop, rupture dimensions and average slip as a function of the rupture velocity.** The constant stress drop estimated using the scaling of Figure 2 is shown (coloured circles) for a rupture velocity that varies from 0.02 $\beta$ (S-wave velocity) to 0.9 $\beta$ (see Methods; Supplementary Table 1). The minimum and maximum rupture dimensions of the LFEs are shown (circle labels), as are their average slips (circle colours). The sizes of the circles are scaled to the average rupture dimensions. Tidal stresses[56] are shown (grey box). Reference values for the rupture velocity are given at the top of the Figure, assuming $\beta = 3.7$ km s$^{-1}$.

We derived a very low stress drop in the kilopascal range only under the assumption of a fast rupture speed (Fig. 3), as observed for regular earthquakes and consistent with previous studies[54,55,29,40], and an average slip of the order of tens of micrometres over sub-kilometre rupture sizes. Assuming a value of 2% of $\beta$ as the lower end member for $v_R$, we derive a stress drop of the order of $10^6$ Pa, with an average slip of the order of 10 cm over rupture sizes that vary from 20 m to 100 m. Thus, the scaling does not constrain the size, slip and stress drop of the LFE sources, as long as independent estimations of the rupture velocity or the rupture size are not available.

At this stage, the interpretation of the results of this study opens up challenging questions.
A rupture speed close to the asymptotic limit leads to a stress drop lower than expected for tremor modulation by tides and surface waves of large teleseismic earthquakes[56,54]; also, it is difficult to reconcile a micrometre scale for the slip with a wide kilometre-scale rupture that occurs along a heterogeneous plate interface.
On the other hand, extremely low rupture speeds are associated with a stress drop up to $10^6$ Pa. Slip and stress drop in this domain lead to an energy budget of the same order as for regular earthquakes. Thus, a completely different and highly dissipative rupture dynamics is required to limit the effective rupture speed at such a small fraction of the shear-wave speed. These models should also explain how such a stress scale of ~$10^6$ Pa can be so sensitive to much smaller stress changes.
Thus, in conclusion, the central region of Figure 3 seems the most likely. A rupture velocity smaller than $0.4\beta$ is compatible with previous modelling of apparent LFE source time functions[57], and down to velocities ~$0.05\beta$ the related stress drop scale (1-100 kPa) is consistent with modulation by tides and teleseisms; the averaged slip varies from 0.1 mm to 1.0 mm over averaged rupture sizes from 400 m to 100 m.

The mean value of our scaling exponent estimate is slightly smaller than -3; this might implicate that the stress drop weakly scales with the rupture dimension. Nevertheless, the stress drop variation would be within one order of magnitude. At this stage, uncertainty in the estimations do not allow to discriminate between this behaviour and the regular one, and we chose to discuss the latter, simpler, interpretation.

**Acknowledgements**

We thank N. M. Shapiro, S. Ide, T. Uchide, P. Bernard and J. Renou for really helpful discussions that lead to significantly improve the manuscript.

This study was supported by the European Research Council under the European Union Horizon 2020 research and innovation program (grant agreement no. 787399 - SEISMAZE). Most numerical computations were performed on the S-CAPAD platform, at Institut de Physique du Globe de Paris (IPGP), France.

**Author contributions**

M.S. analysed the seismic data and estimated the source parameters, converted hypocentre and arrival time data from JMA unified earthquake catalogue to provide event waveforms for the LFEs detected by JMA and, with G.F., analysed the dependence of the stress drop on rupture velocity. N.P. provided the main LFE catalogue and event waveforms used in this study. All of the authors contributed to the interpretation of the data, discussions of the results, and preparation of the manuscript.

**Competing interests**

The authors declare that they have no competing interests.

**Methods**

**Extracting low-frequency earthquakes from tectonic tremor sequences**

The massive catalogue of LFEs (Fig. 1; Supplementary Fig. 1) was extracted with the automatic network-based detection and location method BackTackBB[28,34], using 25 Hi-net stations in western Shikoku (Japan) that recorded continuous seismic signals associated with tectonic tremor sequences that occurred from May to June 2012 and January 2014 to November 2016. The method exploits frequency-dependent higher-order statistical signal characteristics to extract and localise in time the onset of short-duration LFE transients within the continuous seismic signals, and uses their coherency across the seismic network to locate the LFE sources in space and time. The methodological processing and analysis steps, together with the set-up parameters, are detailed in Poiata et al.[34].

**Source parameters estimation**

We used a probabilistic method[39] based on the conjunction of states of information between the data and the model to retrieve the LFE source parameters from the joint PDF expected over the model space when both model and data uncertainties are assumed to be normally distributed. We model the S-wave far-field amplitude displacement spectrum of Equation (3),

$$\tilde{u}(f) = \tilde{S}(f)\tilde{G}(f) \qquad (3),$$

where $f$ is the frequency, $\tilde{S}(f)$ is the modulus of the Fourier transform of the source–time function, and $\tilde{G}(f)$ is the modulus of the Fourier transform of the Green's propagator.

The source spectrum is modelled assuming a generalised Brune's spectral model[37], as in Equation (4):

$$\tilde{S}(M_0, f_c, \gamma; f) = M_0/(1 + (f/f_c)^\gamma) \qquad (4).$$

The model space is defined by three source parameters: the seismic moment, $M_0$, which is related to the energy released by the source; the corner frequency, $f_c$, which is a proxy for the rupture length; and the high-frequency fall-off exponent $\gamma$.

The Green's propagator $\tilde{G}(f)$ is assumed to have a frequency-independent attenuation quality factor $Q^{58}$, which was fixed at 300, as provided in the literature[38].

*Signal processing*

We applied the following methodology to each single station S-wave displacement spectrum. The S-wave arrival times $T_S$ are theoretically obtained from the one-dimensional layered velocity model of Kubo et al.[38]. A 4 s S-wave time window was selected (Eq. (5)), together with a noise time-window of the same duration (Eq. (6)):

$$\Delta T_S = [T_S - 1s, T_S + 3s] \qquad (5),$$

$$\Delta T_N = [T_0 - 4s, T_0] \qquad (6),$$

where $T_0$ is the origin time of the event.

The raw signal was processed to remove the instrumental response together with both the constant and linear trends; Hann-function tapering was applied to the first and last 5% of the signal. The signal and noise amplitude spectra were derived by applying fast Fourier transform to the pre-processed signal and noise time windows, respectively. Finally, each spectrum was smoothed in a logarithmic scale using a five-point moving average filter.

For each LFE and each station, the geometrical mean of the smoothed spectra of the two horizontal components was inverted[59].

*Single-station solution*

The LFE signals are characterised by very low signal-to-noise ratios, which are usually a little larger than 1 (Supplementary Fig. 2a). Even when the S-wave train emerges in the time domain, its amplitude is of the same order of magnitude as the noise amplitude, which can affect the spectral shape. Nevertheless, we can observe a region in the frequency domain around the LFE corner frequency where the S-wave spectrum is actually larger than the noise spectrum. This sub-domain is usually large enough to resolve the low-frequency flat level and the high-frequency fall-off decay in the S-wave spectrum (Supplementary Fig. 2b).

The spectral modelling is restricted to the frequency sub-domain where the signal amplitude is at least 1.25-fold the noise. In the example in Figure 2b, this region corresponds to the interval [0.8 - 5.2] Hz, which is indicated by the black horizontal arrows. We invert the displacement spectrum in the selected frequency band to retrieve the joint PDF for the source parameters, with the estimation of the expected value and related uncertainty for each parameter as the mean and the standard deviation of the corresponding marginal PDF (Supplementary Fig. 2c). In Supplementary Fig. 2b, we show an example of the theoretical spectrum, as calculated with the estimated source parameters, and superimposed on the observed spectrum. We also show in Supplementary Fig. 2d the 2-D marginal PDFs for each couple of parameters.

*Quality selection criteria*

We automatically discard noisy data for which the selected frequency sub-domain where the S-wave spectrum is above the noise spectrum is reduced to less than 10 points (90% of rejections). Moreover, for some records, the signal-to-noise ratio can be too low, which leads to an unconstrained PDF in terms of at least one parameter. This allows automatic detection and discarding of these unconstrained solutions[39] (10% of rejections).

Application of these two criteria resulted in rejection of about 75% of the LFE events in the catalogue (Supplementary Fig. 1).

*Event solution*

Source parameters for an event are obtained as the weighted means of single-station estimations, where the weights are the inverse of the variances, and their uncertainties are given by the standard errors[39].

*Corner frequency–seismic moment scaling*

We bin the event solutions in the seismic moment domain (Fig. 2) to estimate the scaling coefficient between the corner frequency and the seismic moment. The size of each bin was 0.03. The seismic moment estimation corresponds to the centre of the bin, while the corner frequency is the weighted mean of the event solutions belonging to the bin. The variability of the corner frequency measurements in the bin is represented by the weighted standard deviation shown in Equation (7);

$$\bar{\sigma}_W = \sqrt{\sum_{i=1}^{N}\left[\left(f_{c_i} - \bar{f_c}\right)^2 / \sigma_{f_{c_i}}^2\right] / \sum_{i=1}^{N} \sigma_{f_{c_i}}^{-2}} \qquad (7),$$

where N is the total number of event solutions in the bin (Fig. 2).

In the spectral inversion, we do not consider site effects[39]. The average of the corner frequencies in each bin comes from a large number of stations and mitigates possible single-station site effects, if any.

**Probability density function of the scaling exponent**

We estimated the PDF of the scaling exponent (Fig. 2a, Supplementary Fig. 6) through a Bootstrap method. We randomly extract a single value of $f_c$ per seismic moment bin from a normal distribution parameterised by the mean and variance of the bin (Fig. 2a). This provides a new collection of corner frequencies as a function of the seismic moment. For this set of couples ($M_0$, $f_c$), the scaling exponent is estimated, as discussed in the main text. This extraction procedure is repeated 100,000 times to obtain a good approximation of the PDF of the scaling exponent from the normalised histogram of the estimated scaling exponents.

We estimated the PDF for 3 different seismic moment domains, to assess the robustness of the results. First, we used all the available observations (Fig. 2a); then, we reduced the domain to a decade, selecting the 75% of the observations (Supplementary Fig. 6a). Finally, we reduced the seismic moment domain to half a decade, selecting the 44% of the observations (Supplementary Fig. 6b).

**Dependence of the stress drop on the rupture velocity**

The corner frequency $f_c$ is a proxy for the rupture size $r$. Assuming a self-similar circular fault model with a constant rupture velocity $v_R$ expressed as a fraction of the S-wave velocity $\beta$, the corner frequency can be expressed as a function of the shear-wave speed and the source size $r$ as $f_c = k\beta/r$, where $k$ is a constant that depends on the rupture speed[53,60,61].

We analyse the dependence of the rupture size, and therefore of the stress drop $\Delta\sigma \propto M_0/r^3$ [52,62], on the rupture velocity. For the sake of simplicity, we used the kinematic circular crack model of Sato and Hirasawa[53] to estimate the $k$ coefficient for different $v_R$. This model has an analytical representation of the expected far-field displacement spectrum that allows the computing of the synthetic spectra for different rupture speeds and different take-off angles, from which we estimate the corner frequencies using the probabilistic method of Supino et al.[39].

We estimate the $k$ coefficient[53], averaging the different corner frequencies obtained for a take-off angle $\theta$ from 0° to 90°, with a discretisation step of 5°, to remove the expected directivity effects (Supplementary Fig. 8). Sato and Hirasawa[53] provided $k$-values for $v_R$ ranging from $0.5\,\beta$ (k = 0.25) to $0.9\,\beta$ (k = 0.32). We retrieved the same coefficient values, and extend these estimates to lower rupture velocities. We obtain $k = 0.214$ for $v_R = 0.4\,\beta$, $k = 0.096$ for $v_R = 0.1\,\beta$, $k = 0.061$ for $v_R = 0.05\,\beta$, and $k = 0.028$ for $v_R = 0.02\,\beta$.

The estimated source radius decreases as $v_R$ decreases (Fig. 3), which lead to an increasing stress drop $\Delta\sigma \propto M_0/r^3$. The rupture velocity dependence of $k$ might vary when using different self-similar source models, and other source geometries. However, this will not change the order of magnitude of the estimated rupture size and stress drop; e.g., when $v_R = 0.9\ \beta$, $k = 0.21$ for the Madariaga circular crack model[60], and $k = 0.26$ for the Kaneko and Shearer frictional circular model[61]. Different geometries would also affect the $k$ values[63].

**Supplementary information**

| $A$ (slope)      | $1/A$      | $B$ (intercept) |
|------------------|------------|-----------------|
| $-0.287 \pm 0.008$ | $-3.5 \pm 0.1$ | $3.65 \pm 0.09$   |
| $-1/3$ (fixed)   | $-3.0$     | $4.179 \pm 0.005$ |

**Table S1.** Best fit parameters of the linear regression $\log f_c = A \log M_0 + B$, using the bin-averaged corner frequencies (Fig. 2). The intercept $B$ is also estimated fixing the slope to the value $-1/3$; this estimate is used to compute the values of stress drop shown in Figure 3.

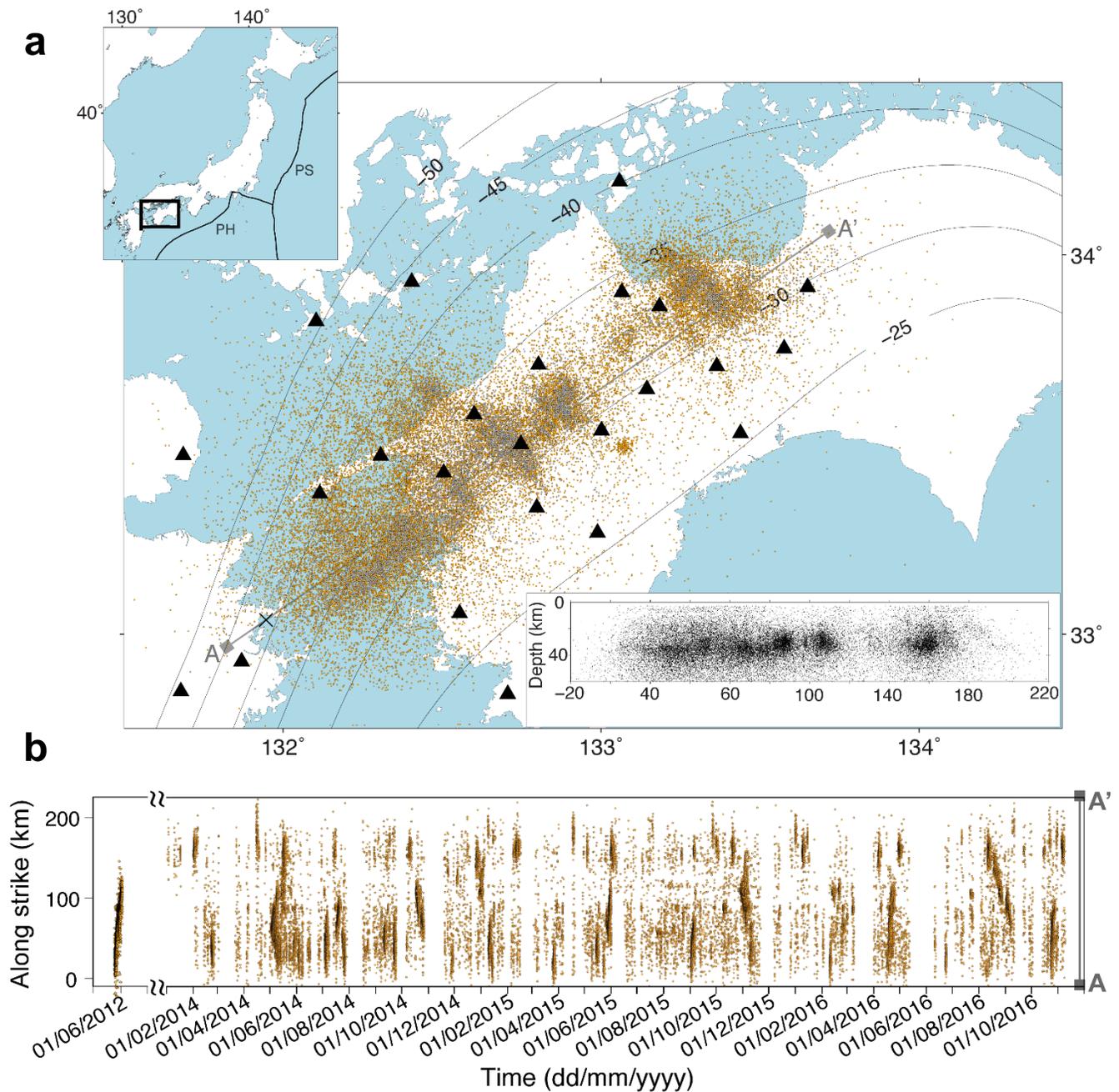

**Figure S1. Distribution of detected low-frequency earthquakes. a**, Map of the locations of the detected low-frequency earthquakes (brown circles) and the Hi-net stations (triangles). The events for which we estimated a source parameters solution are shown (grey circles) (see Methods). Inset, top: The geographic and tectonic settings of the western Shikoku area. Inset, bottom: The depth cross-section of events projected along the strike of N 40º E (A-A'). **b**, Space–time plot of the events. Colours as in (a).

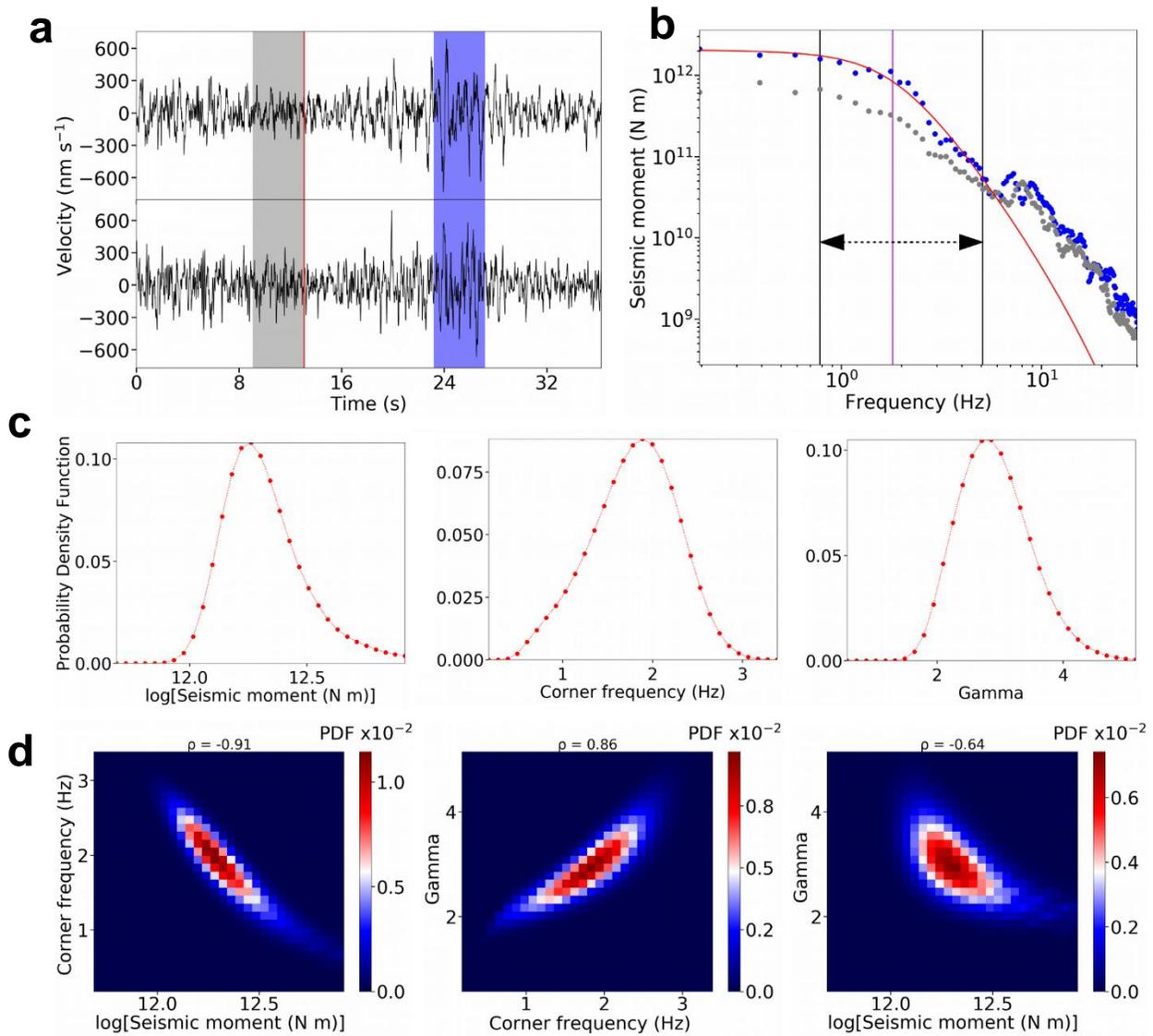

**Figure S2. Source parameter solutions for single-station observation. a**, Selected S-wave signal (blue box) and noise (grey box). The origin time of the event is shown (red bar). **b**, Displacement amplitude spectrum of the unfiltered S-wave signal (blue circles), the noise (grey circles), and the best-fit solution (red curve). The black arrows show the frequency domain selected for the inversion (see Methods). The vertical line shows the estimated corner frequency (magenta). **c**, Marginal probability density functions of the source parameters log $M_0$, $f_c$ and $\gamma$. **d**, 2-D marginal probability density functions of the source parameters; correlation coefficients are at the top of each heatmap. Event-ID 20140505_2358H, Hi-net station N.UWAH.

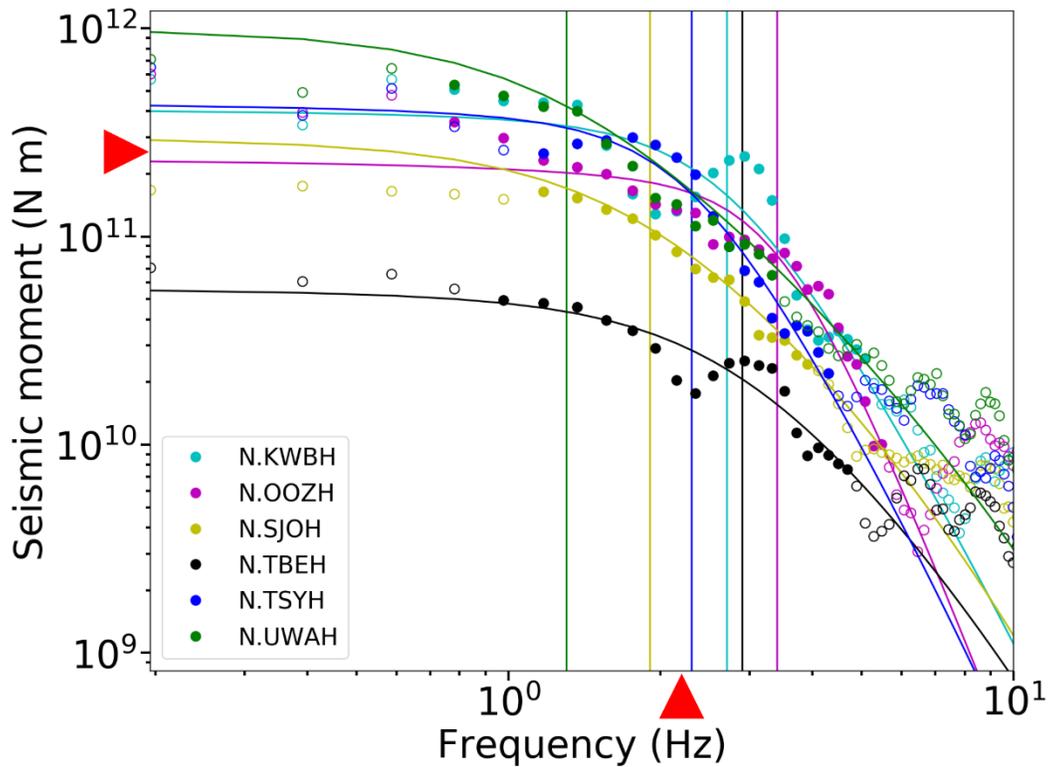

**Figure S3. Seismic moment and corner frequency variability for single-event solutions**. The single-station displacement spectra for which source parameters solutions were retrieved are shown (discrete curves), along with the best-fit solutions (continuous curves) and corner frequency estimates (vertical bars), the frequencies not selected for the inversion due to low signal-to-noise ratio (empty circles) (see Methods), and the corner frequency and seismic moment estimates for the event (red arrowheads). Event-ID 20150530_1157H, Hi-net stations, see Key.

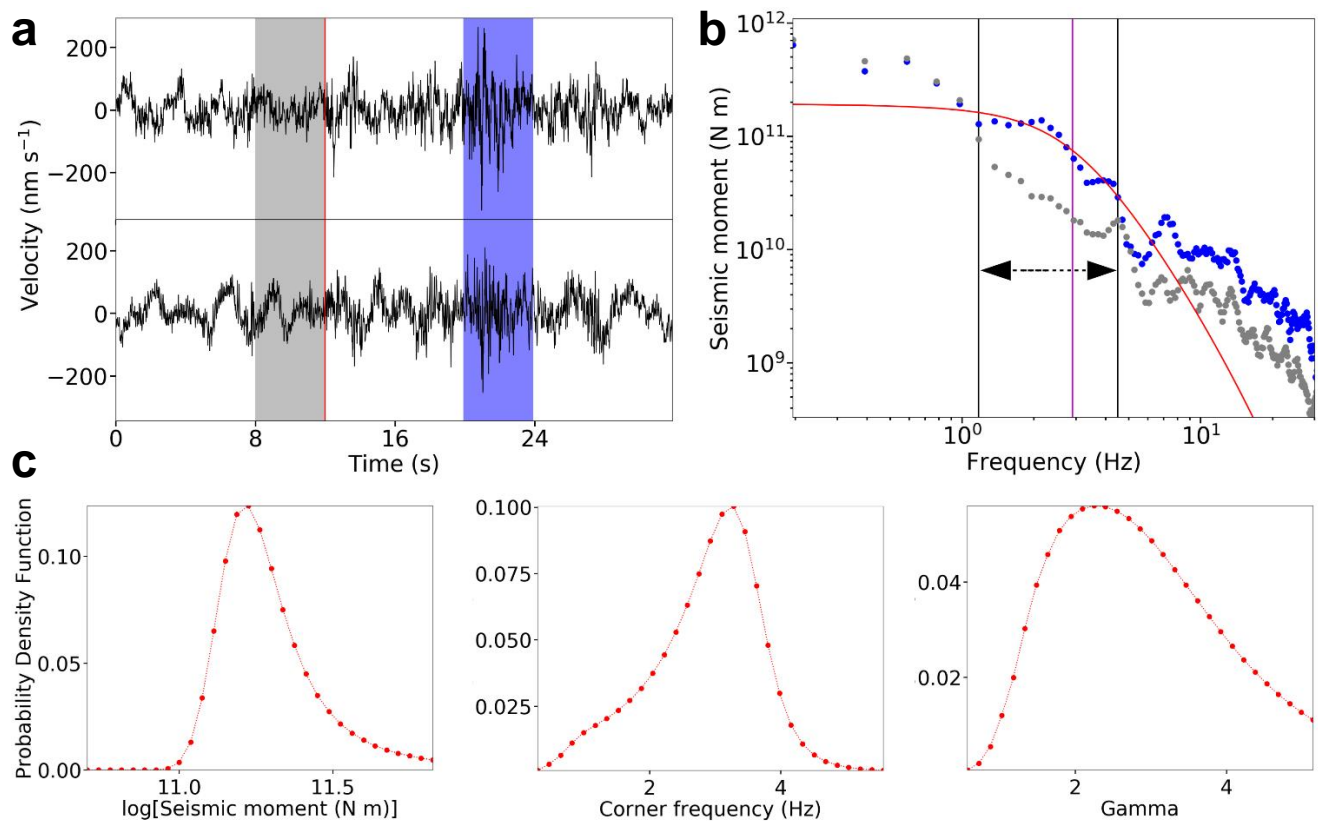

**Figure S4. Source parameter solutions for single-station observation with log $M_0$ = 11.3. a-c,** As for Supplementary Figure 2. Event-ID 20150212_0123Q, Hi-net station N.GHKH.

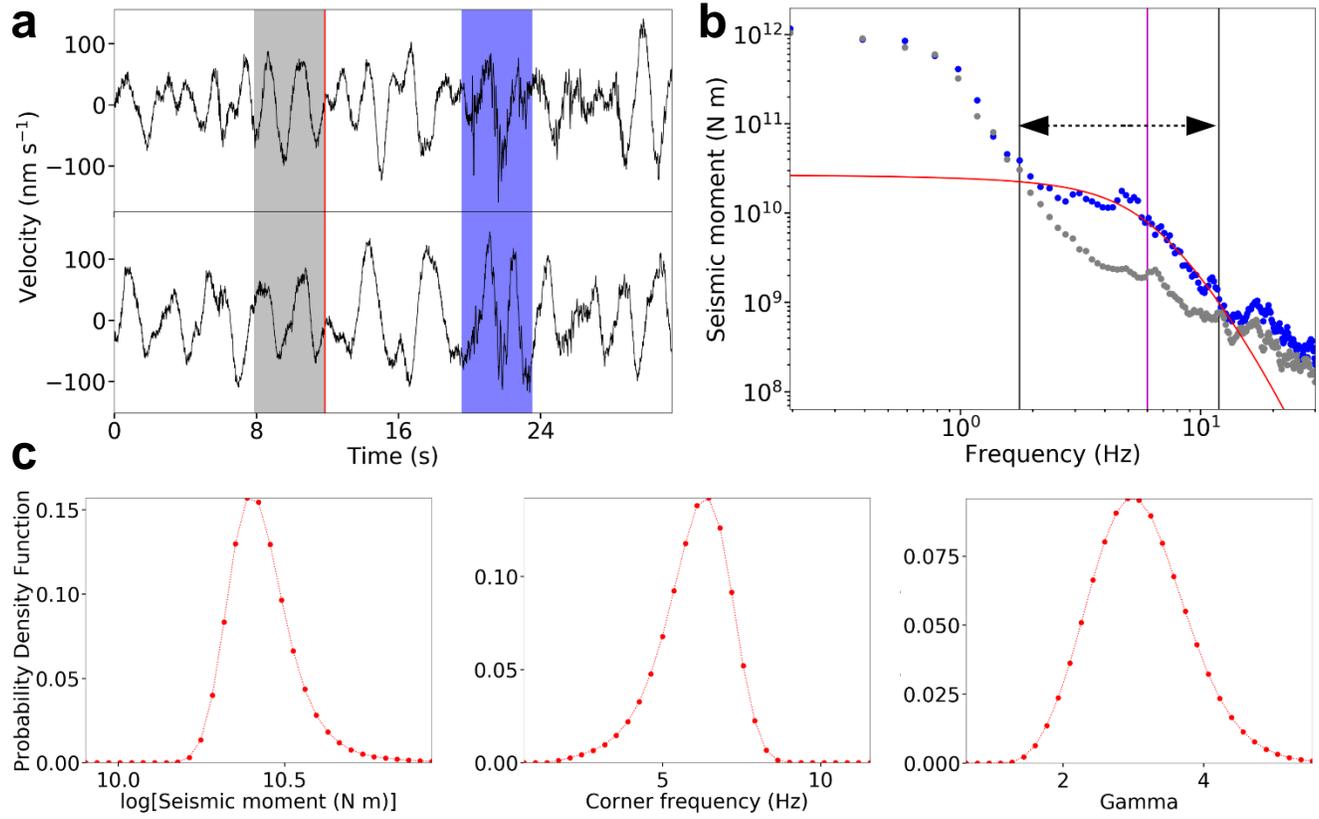

**Figure S5. Source parameter solutions for single-station observation with log $M_0$ = 10.4. a-c,** As for Supplementary Figure 2. Event-ID 20151108_0620B, Hi-net station N.KWBH.

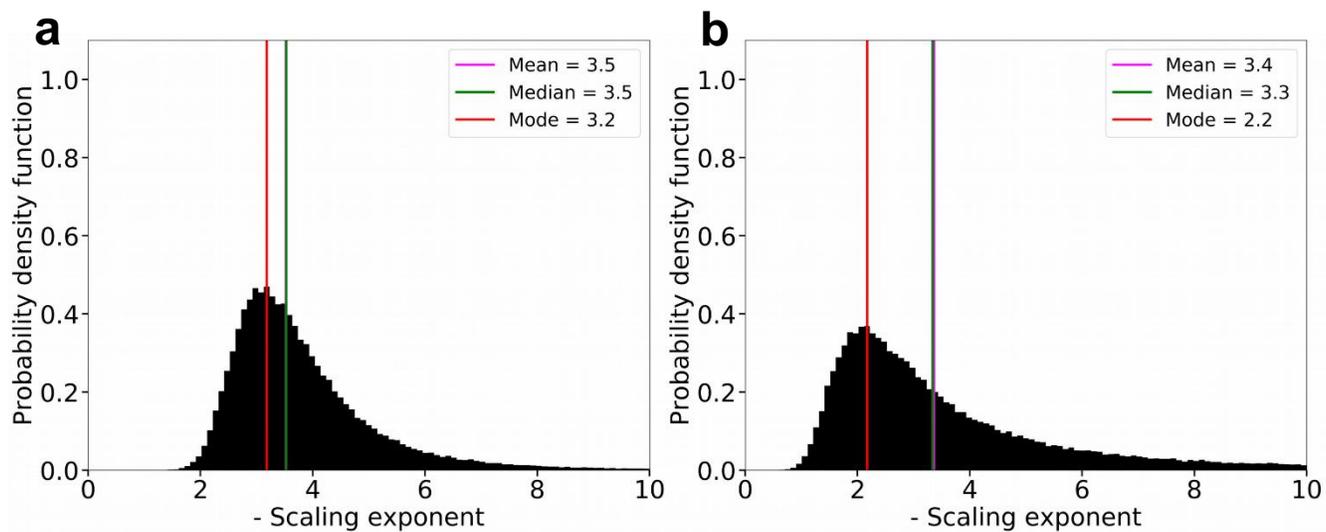

**Figure S6. PDF of the scaling exponent.** Probability density functions of the scaling exponent estimated with a bootstrap method performed with 100,000 random extractions (see Methods). **a**, Bootstrap performed using half of the seismic moment domain shown in Figure 2 (log $M_0$ = 11.0 – 12.0). **b**, Seismic moment domain used for bootstrap test is reduced to half a decade (log $M_0$ = 11.0 – 11.5).

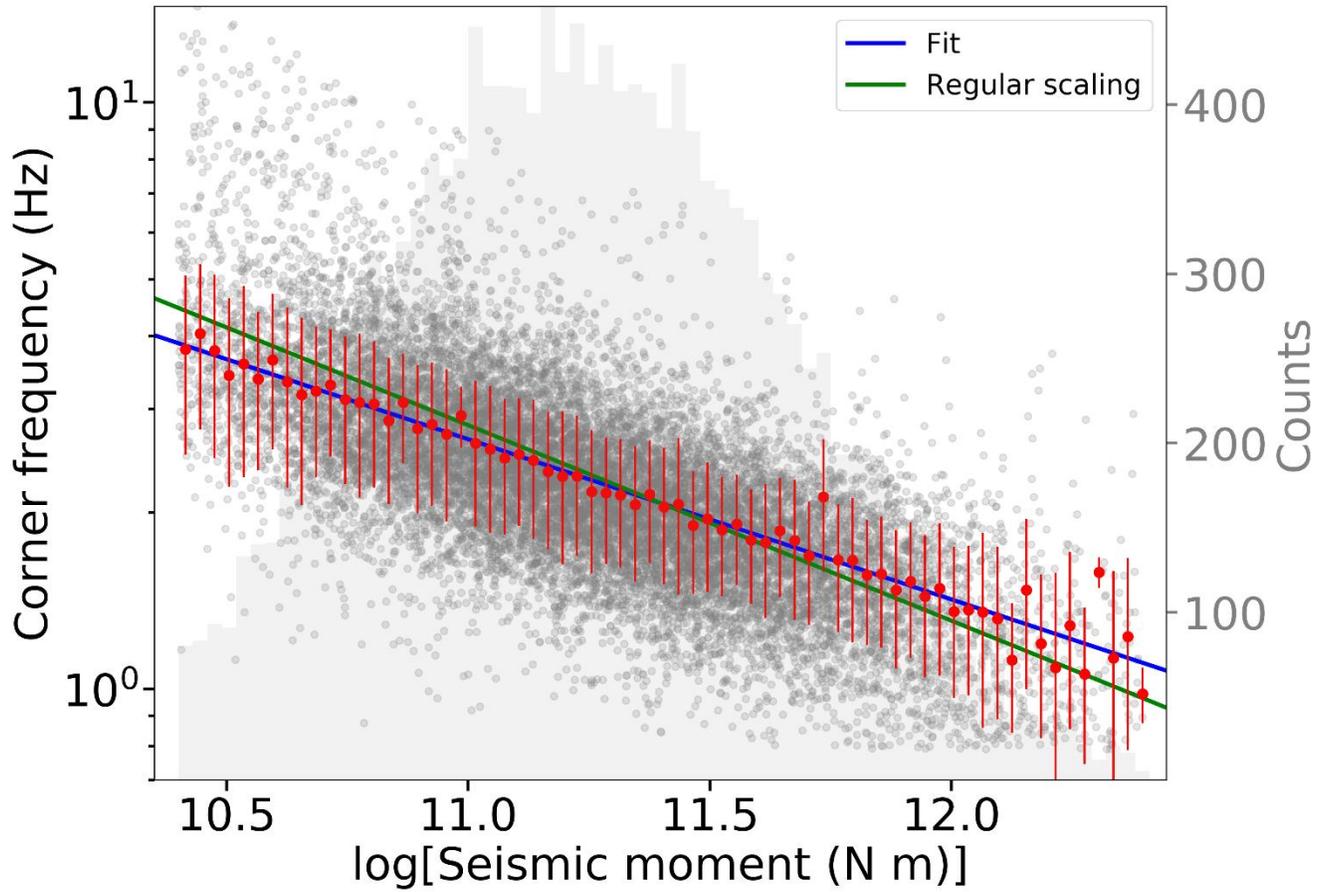

**Figure S7 | Scaling of the corner frequency with the seismic moment, assuming a frequency-dependent Q.** As for Figure 2. The best-fit curve (blue line) of the averaged estimates (red points) has a scaling exponent of -3.7.

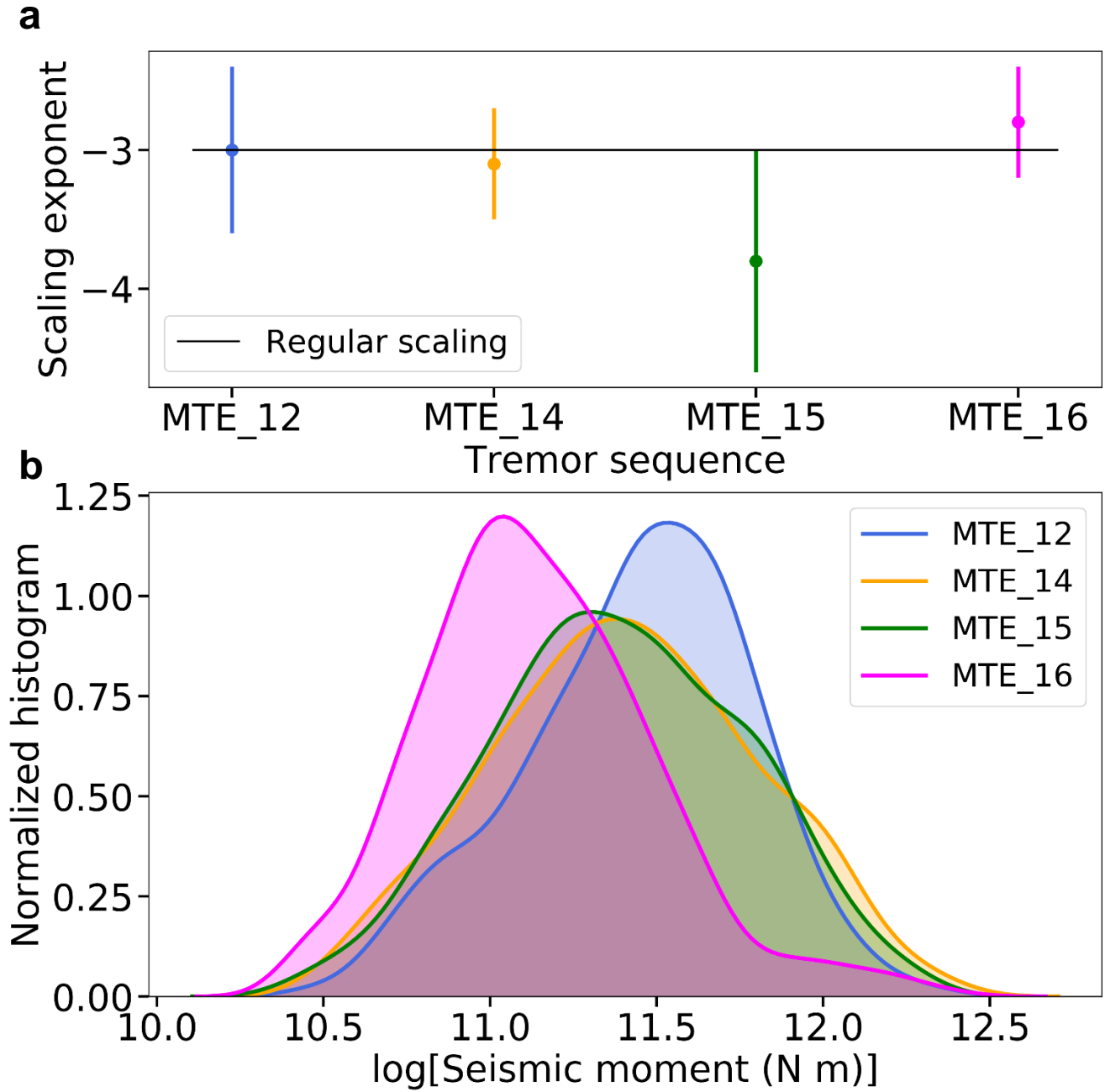

**Figure S8. Scaling exponents for low-frequency earthquake clusters. a**, The scaling exponent is estimated as shown in Figure 2, for different clusters of low-frequency earthquakes. MTE_12, MTE_14, MTE_15 and MTE_16 refer to the major tremor sequences of 2012, 2014, 2015 and 2016, respectively. Error bars: 2-σ standard error. **b**, Normalised histograms showing the explored seismic moment domain for each cluster, with colours as for the top panel. The distributions are similar between the clusters.

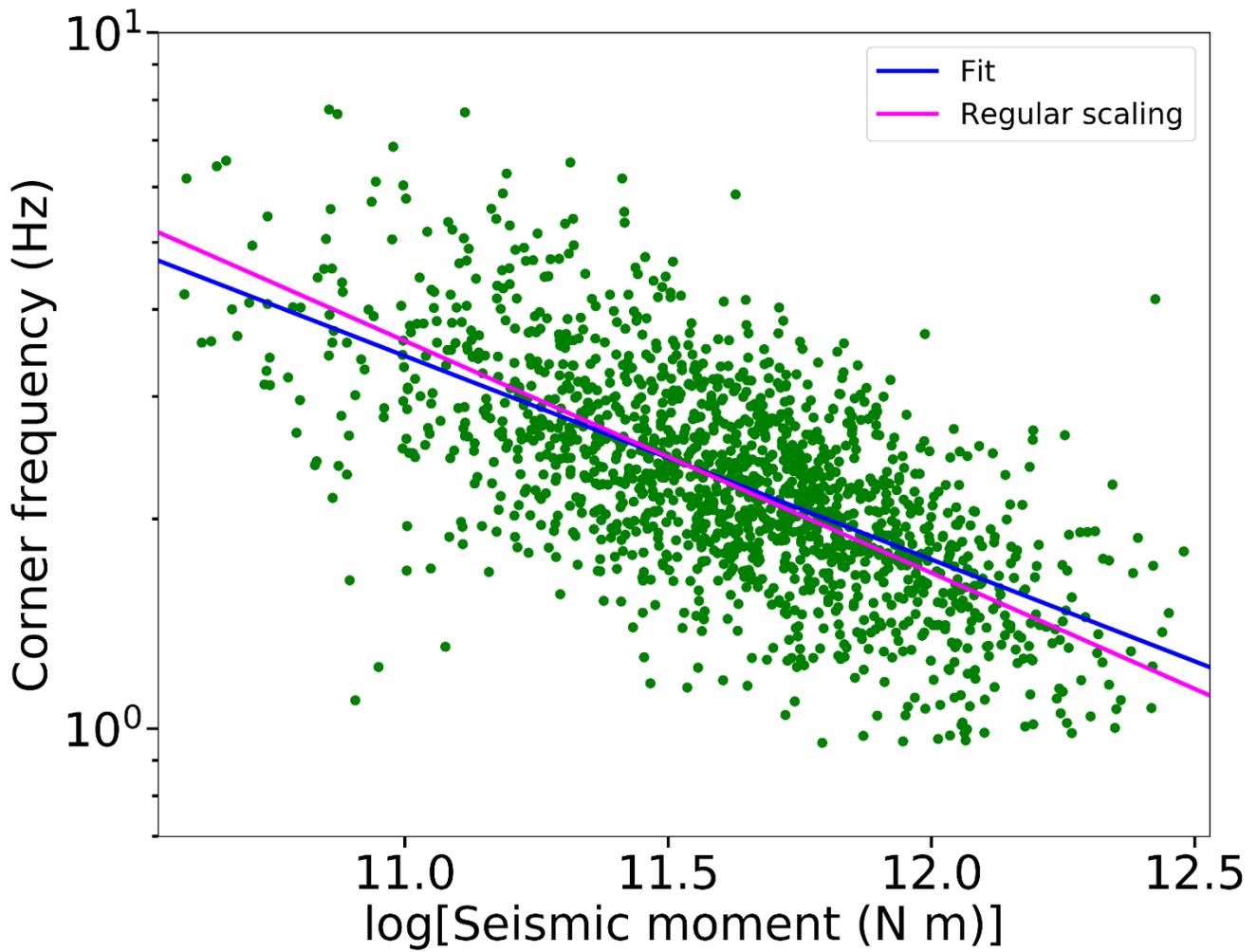

**Figure S9 | Scaling of the corner frequency with the seismic moment, for the JMA catalogue events.** The best-fit curve (blue line) of the estimates (green points) has a scaling exponent of -3.4. The corner frequencies and seismic moment estimates for each LFE are shown (green points). The magenta line represents the regular scaling.

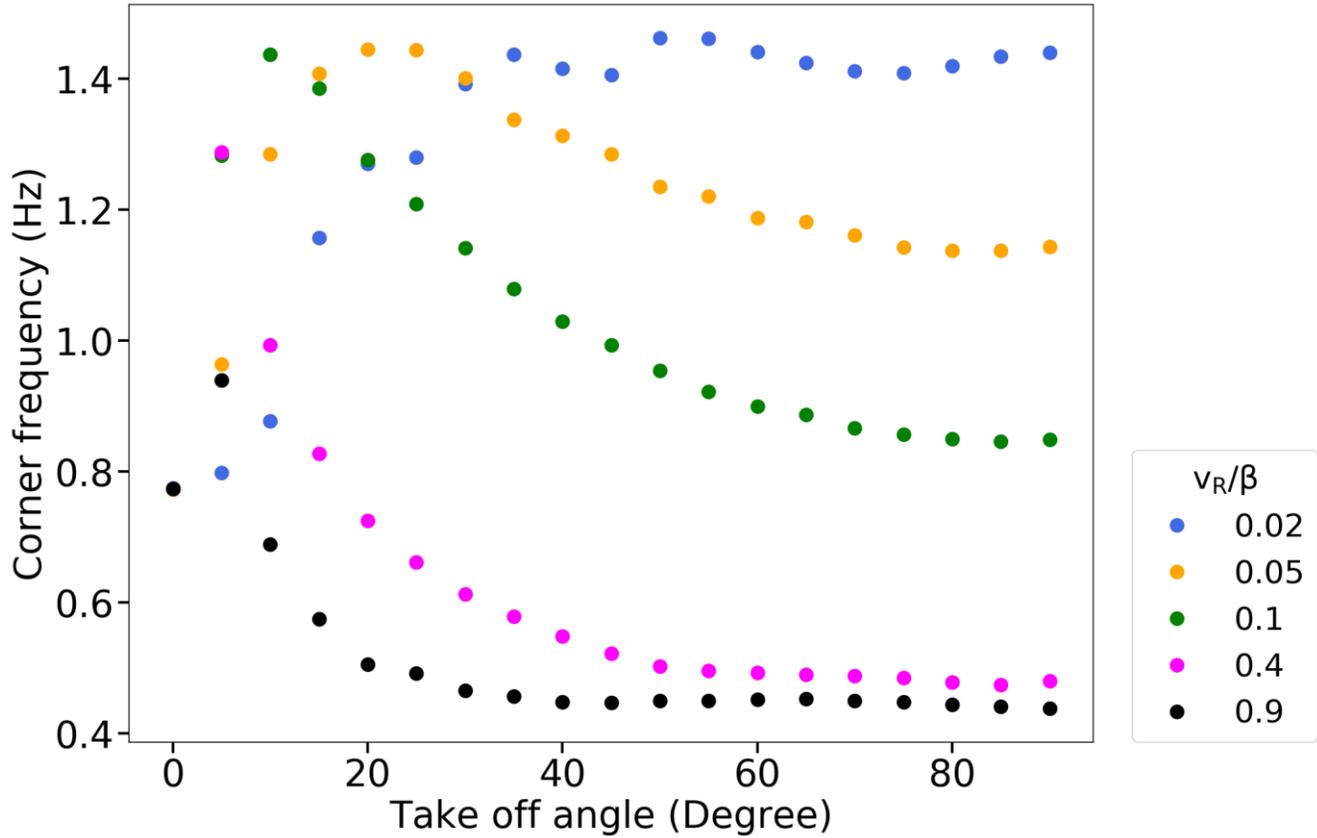

**Figure S10. Corner frequencies as a function of take-off angles for different rupture velocities.** The corner frequencies are estimated from synthetic spectra generated using the Sato and Hirasawa model[46], for take-off angles from 0° to 90°, with discretisation step of 5° (see Methods). Colours referred to the different $v_R/\beta$ used to generate the spectra. $v_R$, rupture velocity; $\beta$, shear-wave velocity, $v_R/\beta$, see Key.